\documentclass[reprint,aps,prl,superscriptaddress]{revtex4-2}
\usepackage{graphicx}
\usepackage{dcolumn}
\usepackage{bm}
\usepackage[usenames]{color}
\usepackage{xcolor}
\usepackage{float}
\usepackage{tabularx}
\usepackage{booktabs}
\usepackage{hyperref}
\usepackage{siunitx}
\usepackage[utf8]{inputenc}
\usepackage[english]{babel}
\hypersetup{colorlinks,linkcolor=red,citecolor=blue,urlcolor=blue,final}
\usepackage{amsmath, amsthm, amscd, amssymb}

\begin{document}
\title{Accelerated State Expansion of a Nanoparticle in a Dark Inverted Potential}

\author{Grégoire F. M. Tomassi}
\affiliation{Nanophotonic Systems Laboratory, Department of Mechanical and Process Engineering, ETH Zurich, 8092 Zurich, Switzerland}
\affiliation{Quantum Center, ETH Zurich, 8083 Zurich, Switzerland}

\author{Dani{\"e}l Veldhuizen}
\affiliation{Nanophotonic Systems Laboratory, Department of Mechanical and Process Engineering, ETH Zurich, 8092 Zurich, Switzerland}
\affiliation{Quantum Center, ETH Zurich, 8083 Zurich, Switzerland}

\author{Bruno Melo}
\affiliation{Nanophotonic Systems Laboratory, Department of Mechanical and Process Engineering, ETH Zurich, 8092 Zurich, Switzerland}
\affiliation{Quantum Center, ETH Zurich, 8083 Zurich, Switzerland}

\author{Davide Candoli}
\affiliation{ICFO – Institut de Ciencies Fotoniques, Mediterranean Technology Park, 08860 Castelldefels (Barcelona), Spain}

\author{Andreu Riera-Campeny}
\affiliation{Institute for Quantum Optics and Quantum Information of the Austrian Academy of Sciences, A-6020 Innsbruck, Austria}
\affiliation{ICFO – Institut de Ciencies Fotoniques, Mediterranean Technology Park, 08860 Castelldefels (Barcelona), Spain}

\author{Oriol Romero-Isart}
\affiliation{ICFO – Institut de Ciencies Fotoniques, Mediterranean Technology Park, 08860 Castelldefels (Barcelona), Spain}
\affiliation{ICREA – Instituci\'o Catalana de Recerca i Estudis Avan\c cats, Barcelona 08010, Spain}

\author{Nadine Meyer}
\email{nmeyer@ethz.ch}
\affiliation{Nanophotonic Systems Laboratory, Department of Mechanical and Process Engineering, ETH Zurich, 8092 Zurich, Switzerland}
\affiliation{Quantum Center, ETH Zurich, 8083 Zurich, Switzerland}

\author{Romain Quidant}
\affiliation{Nanophotonic Systems Laboratory, Department of Mechanical and Process Engineering, ETH Zurich, 8092 Zurich, Switzerland}
\affiliation{Quantum Center, ETH Zurich, 8083 Zurich, Switzerland}

\date{\today}
\begin{abstract} 
While the wave packet of a massive particle grows linearly under free dynamics, it grows exponentially in an inverted harmonic potential, offering a pathway to rapidly increase quantum fluctuations to macroscopic dimensions. In this work, we experimentally demonstrate this principle by expanding the center-of-mass thermal state of a \SI{125}{\nano m} silica nanoparticle to a position uncertainty of \SI{43.4}{\nano m} within \SI{260}{\micro s}. This expansion, achieved using an inverted dark potential to minimize decoherence from photon recoil, represents a 952-fold increase, reaching a scale comparable to the nanoparticle's physical size. This work represents a key advancement toward preparing macroscopic quantum superpositions at unprecedented mass and length scales.
\end{abstract}

\maketitle

Quantum mechanics predicts that a massive particle cannot exist in a motional state exhibiting no uncertainty in phase-space, even at zero temperature. Specifically, the Heisenberg principle states that due to the non-commutation of a pair of canonically conjugated variables, such as position and momentum, the product of their standard deviations is bounded from below by $\hbar/2$~\cite{heisenberg1927anschaulichen}. These Heisenberg-limited 
states can be prepared by reducing the entropy of their motional state, namely, through cooling. This has been experimentally achieved with massive objects confined in harmonic potentials, including a single ion~\cite{diedrich1989laser,monroe1995resolved,roos1999quantum}, a nanoparticle containing billions of atoms~\cite{delic2020cooling, magrini2021real, tebbenjohanns2021quantum, ranfagni2021vectorial, kamba_optical_2022, piotrowski2022simultaneous}, and various micromechanical resonators~\cite{aspelmeyer2014cavity,oconnell2010quantum,teufel2011sideband,chan2011laser,verhagen2012quantum}, with some resonators having masses as large as $\sim 10^{19}$ atomic mass units~\cite{bild_schrodinger_2023}. In harmonic potentials of frequency $\Omega$, the position standard deviation of these Heisenberg-limited states is on the order of the so-called zero-point motion $\sqrt{\hbar/(2m\Omega)}$, where $m$ is the mass of the object. Consequently, the more massive the object, the smaller the position uncertainty for these Heisenberg-limited states prepared in a harmonic potential. Thus, it 
appears that the more massive an object is, the more microscopic its quantum effects become.

However, quantum mechanics allows, in principle, the amplification of quantum fluctuations by increasing the standard deviation of one canonical variable while reducing the conjugate one,
such that their product remains close to $\hbar/2$. These motionally squeezed states can exhibit position uncertainties much larger than $\sqrt{\hbar/(2m\Omega)}$, potentially extending beyond the size of the particle like in matter-wave experiments~\cite{davisson_scattering_1927,keith_diffraction_1988,arndt_waveparticle_1999,arndt2014testing}. Such macroscopic quantum states of massive objects become extremely sensitive to external signals and decoherence, and can be used to test quantum mechanics in regimes where collapse models predict the breakdown of the superposition principle~\cite{romero2011quantum,bassi2013models}. Additionally, they can be 
transformed into non-Gaussian states exhibiting negativities in their Wigner function~\cite{rosiek2024quadrature,riera2023wigner}. 
In clamped oscillators, where the tight harmonic potential is fixed, moderate motional squeezing up to \SI{4.7}{dB}~\cite{lei_quantum_2016} was reported. 
Recently, the possibility of quenching the harmonic potential of a levitated nanoparticle and exploiting motional dynamics in the absence of laser light -- which would generate decoherence due to photon scattering -- has been proposed as a promising platform for achieving unprecedented levels of motional squeezing~\cite{weiss2021large, cosco2021enhanced, roda2024macroscopic,neumeier2024fast}. 
A particularly promising method involves using a dark (non-optical) inverted harmonic potential since, in this case, 
the generation of motional squeezing is exponential in time~\cite{weiss2021large,pino2018chip,romero2017coherent}. This stands in stark contrast to the linear regime induced by dynamics in shallower harmonic potentials or even free dynamics. Exponentially fast expansion of states is relevant to overcome challenges imposed by decoherence 
and stability, both affecting the dynamics during the protocol as well as the uncertainties between experimental runs~\cite{weiss2021large,roda2024macroscopic}. 

In recent years, experiments aiming at the ambitious goal of preparing macroscopic quantum superposition states of nanoparticles have been proposed~\cite{romero2011quantum,romero2011large,scala2013matter,stickler2018probing,pino2018chip,weiss2021large, cosco2021enhanced, roda2024macroscopic, neumeier2024fast}. The often required ground-state cooling of an optically levitated nanoparticle was achieved using either an optical cavity~\cite{delic2020cooling, piotrowski2022simultaneous, pontin_simultaneous_2023, ranfagni_two-dimensional_2022} or active feedback~\cite{tebbenjohanns2021quantum, magrini2021real, kamba_optical_2022, kamba2023nanoscale}.
Furthermore, motional state expansion, linear in time, were reported using either shallower optical potential or free dynamics, starting from both the ground state~\cite{rossi2024quantum,kamba_revealing_2023} and thermal state \cite{Ulbricht,hebestreit2018sensing} of a nanoparticle.
However, squeezing protocols that utilize optical means are limited by decoherence induced by photon scattering~\cite{jain2016direct,chang2010cavity,maurer2023quantum}. Recently, 
hybrid schemes have been reported in the classical regime~\cite{conangla2020extending,bykov2022hybrid,bonvin_hybrid_2024}, where an optically levitated nanoparticle is released into a shallower harmonic electrical potential~\cite{bonvin_state_2024}. 
In this article, we report a significant progress: the controlled release of an optically trapped and cooled nanoparticle into an inverted dark potential, resulting in an exponentially fast expansion of its motional state \cite{romero2017coherent}. This is achieved using a hybrid electro-optical platform that integrates state-of-the-art motion detection and cooling to low phonon occupations with engineered on-chip electrostatic~\cite{melo_vacuum_2024} and radio-frequency (RF) potentials. Our approach enables rapid squeezing protocols while effectively suppressing photon recoil heating.

\begin{figure}
\centering
\includegraphics[width=0.48\textwidth]{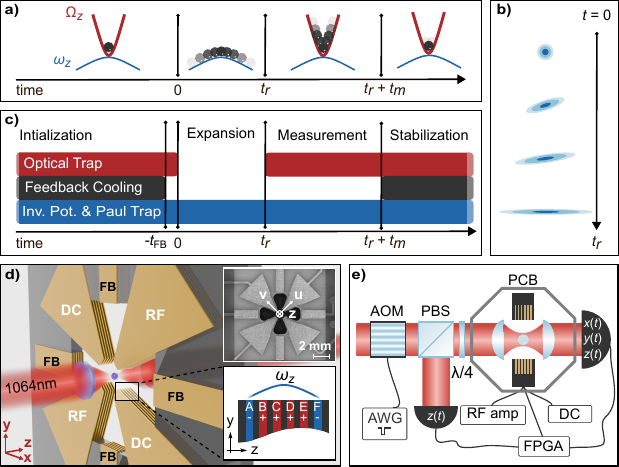} 
\caption{\textbf{State expansion in dark inverted potentials.} 
\textbf{a)} 1D particle trajectory during the experimental protocol. The particle is released from a tight optical trap at $\Omega_z$ (red) into an inverted, electrostatic potential at $\omega_z$ ($t = 0$). The particle delocalizes in the dark and its position at $t_r$ maps into the oscillation amplitude in the optical tweezers before re-initialization ($t_r + t_m$). 
\textbf{b)} Expected phase-space evolution in an inverted potential. The position variance $\sigma_z(t)^2$ in the inverted harmonic potential grows in time until the particle is retrapped at $t=t_r$.
\textbf{c)} Experimental sequence for state expansion in a dark inverted potential. The particle state is released from the optical trap at $t=0$ and re-trapped at $t=t_r$. The feedback is turned off at $-t_\text{FB} = \SI{-5}{\micro \second}$. A lock-in operation using a $t_m = \SI{500}{\micro \second}$ time trace in the optical trap reconstructs the particle state at $t=t_r$. The sequence is repeated 400 times to reconstruct the phase-space distribution. 
\textbf{d)} Hybrid electro-optical chip. A nanoparticle is trapped at the laser focus (wavelength $\lambda =\ $ \SI{1064}{\nano m}, $\text{NA}=0.77$) and surrounded by a 6-layer PCB. The radio-frequency electrodes create Paul trap confinement in the $x,y$-plane ($V_{\text{RF}} =\ $\SI{800}{\volt}, $\Omega_{\text{RF}}/(2\pi) =\ $ \SI{25}{\kilo \hertz}), cold damping FB electrodes cool the nanoparticle CoM  to $\bar{n}_{x,y,z} \approx [721, 3763, 10]$, and DC electrodes generate an electrostatic potential along $z$ ($V_{A,F} =-\SI{14}{\volt}$,$V_{B-E} =+\SI{14}{\volt}$). Top insert: PCB with the Paul trap axes ($u,v$). Bottom insert: voltage configuration for inverted harmonic potential. 
.\textbf{e)} Optical setup. The particle position is measured and feedbacked using homodyne back-detection ($z$) and balanced forward detection ($x,y$). The laser is switched on and off with an acousto-optic modulator (AOM) controlled by an arbitrary waveform generator (AWG). The Paul trap is controlled by an RF amplifier and the feedback signals are processed by a field programmable gate array (FPGA).}
\label{fig:1}
\end{figure}

More specifically, we study the expansion of a nanoparticle's state in a dark, inverted, and electrostatic potential $U(z) = -m \omega_z^2 z^2 /2$ parametrized by the frequency $\omega_z$. The levitated nanoparticle is released from a tighter optical harmonic trap $U_0(z)=m \Omega_z^2 z^2/2$  at the center of the inverted potential, leading to an exponentially accelerated state expansion \cite{romero2017coherent,weiss2021large, roda2024macroscopic}. Throughout the entire protocol described here, the particle occupies a Gaussian state that is fully described by the mean, variances and covariance of the position $\hat{z}$ and momentum $\hat{p}$ quadratures.
 Initially, the particle occupies the symmetric phase-space of a thermal Gaussian state with an average phonon occupation of the center-of-mass motion (CoM) given by the Boltzmann distribution $\Bar{n} = [\exp(\hbar \Omega_z/(k_\mathrm{B} T))-1 ]^{-1}$, before the state expands in phase-space following the inverted harmonic potential, as depicted in Fig.~\ref{fig:1}a-b. 
 In the zeroth order secular approximation~\cite{major2005charged} neglecting micromotion and damping, the position variance $\sigma_z(t)^2 = \langle \hat{z}(t)^2\rangle - \langle \hat{z}(t)\rangle^2$ follows~\cite{graham_squeezing_1987}  
 \begin{equation}\label{eq:vxc}
 \begin{aligned}
    \sigma_z(t)^2 &=& \sigma_z(0)^2 \left[\cosh^2(\omega_z t) +  \frac{\Omega_z^2}{\omega_z^2} \sinh^2(\omega_z t)\right] \\
    & & - \frac{\hbar \Omega_z \Gamma^1_z}{m \omega_z ^2} \left[ t - \frac{\sinh (2\omega_z t)}{2 \omega_z} \right],   
    \end{aligned}
\end{equation}
where the first term describes the coherent dynamics and the second term the incoherent dynamics. The parameter $\Gamma^1_z$ corresponds to the displacement noise and links to the heating rate $\dot{E}_z = \hbar\Omega_z\Gamma^1_z$, which is here dominated by the residual gas and electric field noise (see Supplemental Material Section IX).
The expansion coefficient $\eta_z$ after the release time $t_r$ is defined as the ratio of the initial and final delocalization length $\eta_z = \sigma_z(t_r)/\sigma_z(0)$. 
Note that our setup allows us to compare the frequency jump protocol~\cite{janszky1986squeezing,graham_squeezing_1987,bonvin_hybrid_2024} simultaneously in the transversal directions.

\begin{figure*}
\includegraphics[width=\textwidth]{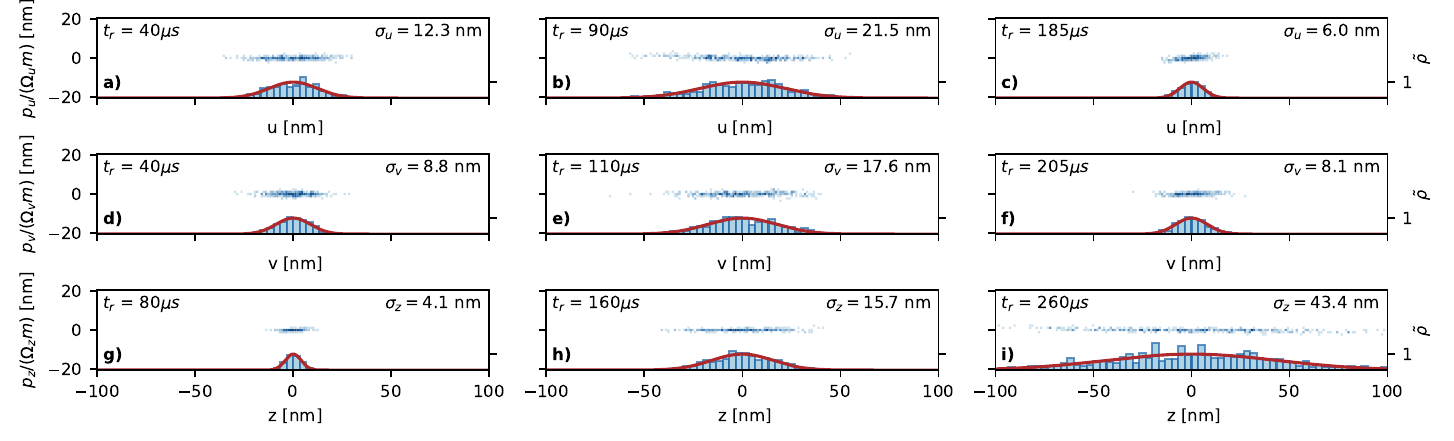}
\caption{\textbf{Phase-space distribution of the particle's CoM} in the $u$- \textbf{(a-c}), $v$- (\textbf{d-f}) and $z$-direction (\textbf{g-i})  at selected $t_r$ expanding in electrostatic potentials. The 1D histograms show the amplitude normalized, projected position distribution $\Tilde{\rho}$. The red curves represent the estimated Gaussian with standard deviation $\sigma_j(t)$, $j \in \{u,v,z\}$. The distributions are constructed from 400 repetitions of the experimental sequence (Fig \ref{fig:1}c) and plot as 2D histograms. The frequencies $\Omega_{u}=\Omega_v=173\text{kHz}$ are defined assuming the $uv$-system is rotated by $\theta_t \approx 45^\circ$ to the $xy$-basis (see Supplemental Material Section IV). The $u$- and $v$-directions are dominated by Paul trap dynamics with maximum position variances $\sigma_{u,v} =[21.5\text{nm},17.6\text{nm}]$  at $t_r = T_{u,v}/4 = \pi/(2\omega_{u,v})$ (\textbf{b, e}), after which $\sigma_{u,v}(t)$ decreases due to the confining potential, resulting in a recompressed state at $t_r = T_{u,v}/2 = \pi/\omega_{u,v}$ (\textbf{c, f}). Along the $z$-direction, the particle's position variance grows exponentially as it evolves in a DC inverted potential, reaching a state size of $\sigma_z = \SI{43.4}{\nano\meter}$.} 

\label{fig:2}
\end{figure*}

 We consider a charged, spherical silica (SiO$_2$) nanoparticle of mass $m \approx \SI{1.95}{\femto\gram}$ corresponding to a radius $R\approx $ \SI{63}{\nano m} that is trapped by an optical tweezer, leading to a 3D harmonic potential with mechanical eigenfrequencies $(\Omega_x, \Omega_y, \Omega_z) /(2\pi) \approx [185, 171, 43.5]$ kHz. The optical trap is interfaced to a 6-layer printed circuit board (PCB), as shown in Fig.~\ref{fig:1}d-e, which integrates a Paul trap~\cite{chen2017atomic,chen_sympathetic_2017,brewer_27mathrm_2019,teller_integrating_2023} for dark confinement (RF) with  segmented virtual ground electrodes (DC) and  electrodes for cold damping (FB)~\cite{poggio2007feedback, conangla2019optimal, tebbenjohanns2019cold, melo_vacuum_2024, Kremer2024}.
The individual PCB layers along the optical propagation axis $z$ allow for the generation of an electrostatic, inverted harmonic potential at the center of the PCB by applying tailored DC voltages to each layers (see lower inset in Fig.~\ref{fig:1}d).    
From an independent measurement, the inverted harmonic potential of $\omega_z/(2\pi) =\ $\SI{1.4} {\kilo \hertz} along the optical axis $z$ is characterized for a positively charged particle with $n_q = 84$ elementary charges (see Supplemental Material Section II).
The secular frequencies in the Paul trap plane are $(\omega_{u},\omega_{v})/(2\pi) \approx [2.7,2.5]~\text{kHz}$ where the $uv$-plane is rotated by $\theta_t\approx 45^\circ$ with respect to the $xy$-plane~\cite{bonvin_hybrid_2024}. Deviations from $\theta_t$ are due to fabrication inaccuracies (see top inset in Fig.~\ref{fig:1}d  and Supplemental Material Section I). 

The particle's CoM motion is detected in forward balanced split detection in $x,y,z$~\cite{jan_phd_thesis} and in backward homodyne detection for $z$~\cite{tebbenjohanns2021quantum}. The latter is used for feedback and data acquisition in the inverted potential.
 We initialize the CoM state by applying 3D active feedback with cold damping which reduces the phonon occupation to $\bar{n}_{x,y,z} \approx [721, 3763, 10]$ 
phonons, at pressure $p_{g}\approx 7.7 \times 10^{-7}$ mbar. 
Motional cooling enables to decouple the CoM modes of the particle, such that we can treat their dynamics independently. Here we primarily focus on the inverted dynamics along the $z$-axis. 

\begin{figure}
\centering
\includegraphics[width=0.5\textwidth]{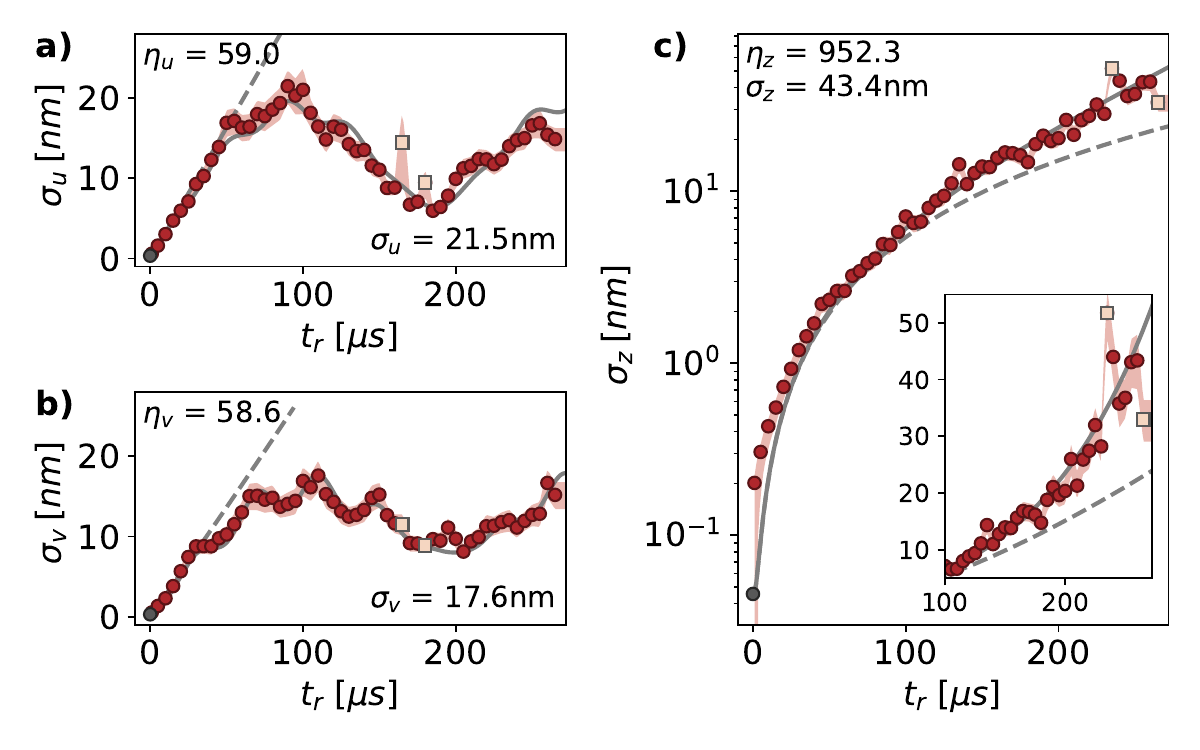}
\caption{\textbf{State expansion of the particle's motional state in a dark electrical potential.} 
\textbf{a-b)} In the frequency jump protocol in the Paul trap $\Omega_{x,y} \rightarrow \omega_{u,v}$,  the particle state uncertainty $\sigma_{u,v}(t_r)$ expands to a maximum at $t_r = T_{u,v}/4$ followed by a recompression at $t_r = T_{u,v}/2$. Maximum state sizes of $\sigma_{u,v}(T/4)=[21.5,17.6]\ \text{nm} $ and expansion ratios of $\eta_{u,v} \approx [59,58.6]$ are reached. 
\textbf{c)} In the inverted DC harmonic potential along $z$, the phase-space expansion is unbound with a maximal state size $\sigma_{z}(t_r = \SI{260}{\micro\second}) = 43.4\text{nm}$ corresponding to $\eta_{z} \approx 952.3$.  
The shaded error bars represent the uncertainty on the position standard deviation and the heating during the measurement.
The solid line fits the theory including coherent micromotion (see Supplemental Material Section VII) to the data $\Gamma^1_j, \Omega_j,\omega_j, \sigma_j(0)^2$ and $\phi_j$ with $j\in {u,v,z}$ as fit parameters. The dashed line corresponds to free expansion as comparison (see main text). The shaded square points highlight states deviating from a Gaussian distribution due to experimental imprecisions (see Supplemental Material Section III).}
\label{fig:3}
\end{figure}

The experimental protocol is illustrated in Fig.~\ref{fig:1}c. The pre-cooled nanoparticle is positioned at 
at the center of the Paul trap by aligning the primary optical trap to minimize micromotion. The initial motional state sizes are  $(\sigma_x(0), \sigma_y(0), \sigma_z(0)) \approx [183, 435, 45.6]\ \text{pm}$.
Following the deactivation of the electrical feedback, the particle is released from the optical trap at a fixed phase $\phi$ of the applied RF Paul trap voltage $V_\mathrm{RF}$. The particle then evolves in the dark electrical potential for a duration $t_r$
before being recaptured in the optical trap.
Once recaptured, the particle's position is measured for $t_m=\SI{500}{\micro \second}$ before the feedback is reactivated to reinitialize the state. This sequence is repeated 400 times for each release duration $t_r$. Throughout the entire protocol, both the Paul trap and the inverted potential remain active. To compensate for time-dependent position shifts caused by micromotion and electrical stray fields, electrical compensation fields are individually adjusted for each release time using the feedback electrodes. 
The particle's position and velocity at $t_r$ are extracted using a lock-in detection method (see Supplemental Material Section III).
Note that the position standard variation increases during the measurement due to heating sources by $(\delta\sigma_x,\delta\sigma_y,\delta\sigma_z) \approx [91, 102, 321]\SI{}{pm}$, corresponding to $\delta \bar{n}_{x,y,z} \approx [178,206,520]$ phonons related to the initial thermal state with $\Omega_x,\Omega_y,\Omega_z$ ( see Supplemental Material Section VI and IX). This effect overestimates the measured state $\sigma_j(t)^2$, which we add to the error bar in Fig.~\ref{fig:3}. 
However, the contribution of $\delta\sigma_j$ to the position uncertainty, overestimating $\sigma_j(t)$, is only significant for small states ($t_r\leq \SI{15}{\micro\second} $).

\paragraph{Results} Our experimental setup enables a direct comparison between the dynamics in an inverted potential along $z$ ($\Omega_z  \rightarrow i\omega_z$) and the dynamics induced by the frequency jump protocol along $u,v$  ($\Omega_{x,y} \rightarrow \omega_{u,v}$)\cite{weiss2021large,  cosco2021enhanced, bonvin_state_2024, rossi2024quantum, Ulbricht, janszky1986squeezing}.
In Fig.~\ref{fig:2} we plot the phase-space distribution for $u,v,z$ at different release times $t_r$ where the phase-space angle due to experimental delays is adjusted in post-processing (see Supplemental Material Section III). The phase-space distributions are represented as 2D histograms constructed from 400 repetitions of the experimental protocol. For each $t_r$, the 1D histogram in the inset shows the projection on the position axis. These are overlaid with a normalized Gaussian curve with the extracted standard deviation $\sigma_j(t)$, visually confirming the Gaussian distribution of the data. In Fig.~\ref{fig:2}a-f, we observe the expansion of the initial state, followed by recompression along $u$ and $v$. In contrast, we observe an ever-growing expansion along the inverted potential in the $z$-direction as can be seen in Fig.~\ref{fig:2}g-i. 
Note that only the major axis of the phase-space distribution can be resolved reliably, unlike the minor axis due to heating effects during the measurement and insufficient accuracy in the phase-space angle (see Supplementary Section III and IX) .

In Fig.~\ref{fig:3} we depict the position standard deviation $\sigma_j(t)$ of the phase-space distribution in dependence of the release time $t_r$ along $u,v,z$. The shaded error bars represent the standard deviation of the position uncertainties and the measurement contribution to $\sigma_j(t)$ associated with $\delta\bar{n}_{x,y,z}$ overestimating the state size.
For the frequency jump protocol along $u,v$ in Fig.~\ref{fig:3}a-b, we observe a maximum expansion at $T_{u,v}/4$ followed by a recompression at $T_{u,v}/2$ where $T_{u,v} =2\pi/\omega_{u,v}$ as predicted by the theory (see Supplemental Material Section VII).  The micromotion due to the Paul trap confinement leads to additional smaller amplitude oscillations dictated by the Paul trap frequency  $\Omega_{\mathrm{RF}}$ (see Supplemental Material Section VIII). 
In Fig.~\ref{fig:3}c we observe the exponentially accelerated state expansion along the inverted potential. The micromotion plays only a negligible role in the inverted potential. 
The solid line is a fit to the theory (Eq.~\ref{eq:vxc} for $z$ and Eq.~46 in the Supplemental Material Section VIII for $u,v$) 
with $\Gamma^1_j, \Omega_j, \omega_j,\sigma_j(0)^2$ and $\phi_j$ as fit parameters (see Tab.~1 in the Supplemental Material). 
The decoherence rate $\Gamma^1_j$ is only considered in the zeroth-order secular motion (second term Eq.~\ref{eq:vxc}), since higher order terms turned out to be negligible. 
For comparison, Fig~\ref{fig:3} displays the state expansion in free dynamics in $j \in \{u,v,z\}$ (dashed line) following
\begin{align}
  \sigma_j(t)^2 &= \sigma_j(0)^2 \left(1 + \Omega_j^2 t^2\right) + \frac{2}{3} \frac{\dot{E}_j}{m} t^3,
\end{align}
where $\dot{E}_j$ is the heating rate accounting for the expansion due to heating.  
From Fig.~\ref{fig:3}, we observe that the free-expansion, the frequency jump and the inverted protocol follow the same dynamics for small release times $t_r$. Nevertheless, the expansion in $u$ and $v$ is bound by the Paul trap confinement to the ratio $\Omega_{x,y}/\omega_{u,v}$, neglecting the micromotion. In contrast, free expansion is unbound but slower than the dynamics in the inverted potential for large $t_r$.
The standard deviation of the state reaches a maximum $\sigma_z=$ \SI{43.4}{\nano m} at $t=$ \SI{260}{\micro s}, corresponding to a state size comparable to the particle's radius and an exponentially fast expansion ratio $\eta_z \approx 952.3$. The final state size, only limited by our linear detection range, beats the free expansion and would continue to grow for longer expansion times, which is the main result of this manuscript.\\

\begin{figure}
\centering
\includegraphics[width=0.5\textwidth]{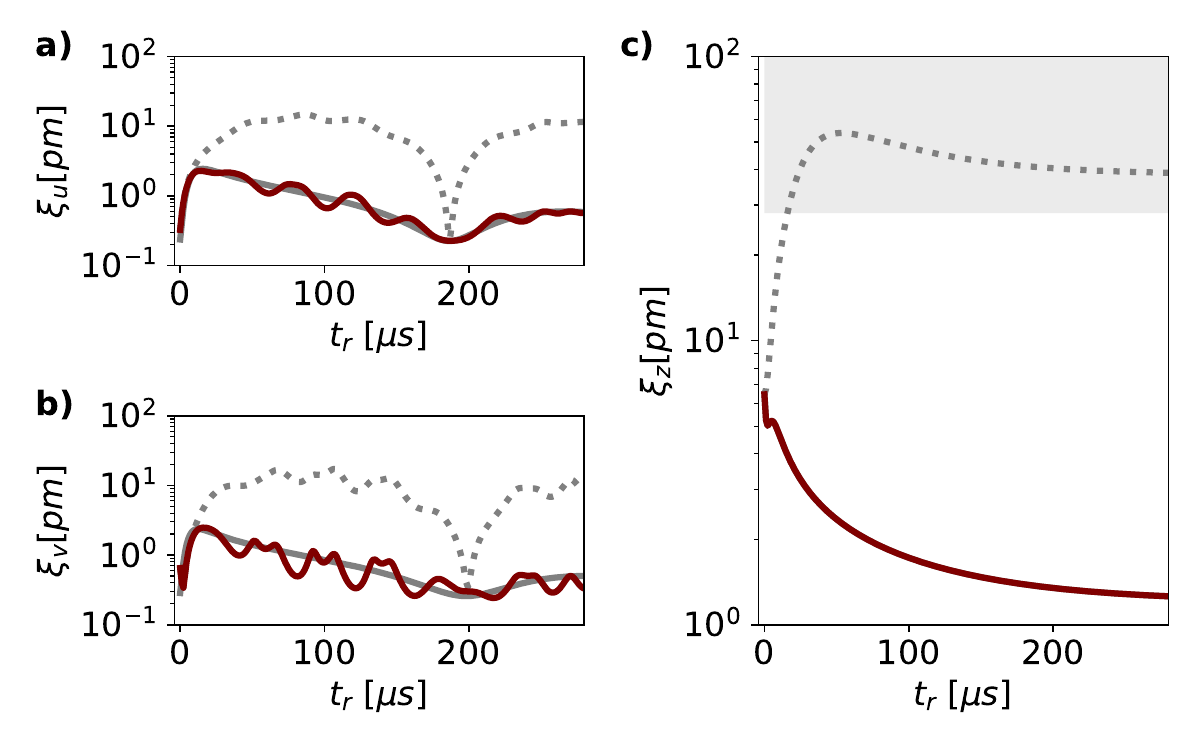}
\caption{\textbf{Evolution of the coherence length $\xi_j(t_r)$ during nanoparticle state expansion} based on fitted parameters $\Gamma^1_j, \Omega_j,\omega_j,\sigma_j(0)^2$ and $\phi_j$ for $j \in {u,v,z}$ related to the data shown in Fig.~\ref{fig:2} and \ref{fig:3}. \textbf{a-b)} The frequency jump protocol shows revivals of $\xi_{u,v}(t)$ with an overlaying decay at a rate given by to $\Gamma^1_j\neq0$. \textbf{c)} Inverted harmonic shows a stabilization of $\xi_z$ after an initial decay. Red solid line: Current heating rate $\dot{E}_j$. Gray dotted line: Improved heating rate $\dot{E}_j/1000$. Gray shaded area indicates $\xi_z(t) > \xi_{|n=0\rangle}$.}
\label{fig:4}
\end{figure}
Nonetheless, to reach large quantum states, it is of utmost importance to expand the state coherently. The figure of merit here is the coherence length \cite{romero2011quantum}
 \begin{equation}
     \xi_j(t) = \sqrt{8} \mathcal{P}_j(t) \sigma_j(t),
 \end{equation}
 with the state purity $\mathcal{P}_j(t) = \mathrm{tr}[\hat{\rho}_j(t)^2] = \hbar/(2\sqrt{\sigma_j(t)^2 \sigma_{p_j}(t)^2 - \sigma_{jp_j}(t)^4})$ where $\bar{\rho}_j(t)$ is the density matrix, $\sigma_{p_j}(t)^2$ the momentum variance and $ \sigma_{jp_j}(t)^2$ the covariance of the Gaussian state. For the case of the optical trap, the purity of the initial thermal state is given by  $\mathcal{P}_j =(2\Bar{n}_j +1)^{-1}$. 
 The variances and covariances depend on the very same parameters, namely $\Gamma^1_j, \Omega_j, \omega_j, \sigma_j(0)^2$ and $m$ (see Supplemental Material VII-VIII).
Under the assumption that our micromotion model captures well the dynamics of the system not only for the measured position variance $\sigma_j(t)^2$ (see Fig.~\ref{fig:3}) but also for the momentum variance $\sigma_{p_j}(t)^2$ and covariance $\sigma_{jp_j}(t)^2$,  we can estimate the coherence length $\xi_j(t)$ by using the fitting parameters from Fig.~\ref{fig:3} (see Tab.~1 in the Supplemental Material).  In Fig.~\ref{fig:4}, we deduce $\xi_j(t)$ for the current experiment (solid red  line) in the presence of a fitted total heating rate of $\dot{E}^{u,v,z}= k_B \times [8.47,11.25,5.91] \text{K/s}$, which stands in reasonable agreement with theoretically predicted values (see Supplemental Material Section IX). Moreover, theory predicts electrical noise to be the largest heating source along  $u,v$, while gas collisions and electric field noise play an equally important role along $z$.  
Starting from $\xi(0)/\xi_{|n=0\rangle} \ll 1$  along $u,v$, we can see an increase of the coherence length for the frequency jump protocol by a factor of $\text{max}[\xi(t)_{u,v}/\xi_{u,v}(0)]\approx [7.1, 3.9]$. 
This gain in coherence length is followed by an oscillatory behavior at the secular frequencies $\omega_{u,v,}$combined with a rapid decay due to $\Gamma^1_j$. Interestingly, smaller oscillations due to micromotion at $\Omega_{\mathrm{RF}}$ can also be observed. In contrast, along $z$, $\xi_z= 6.5$ pm is larger to begin with due to the higher purity of the initial state $\mathcal{P}_z \approx 5\times 10^{-2}$. We observe a partial decay down to $\xi_z(t) = \SI{1.16}{\pico m}$ before $\xi_z(t)$ stabilizes to a nearly constant value, as the expansion nearly compensates for the heating. Notably, reducing $\Gamma^1_j$ by a factor of $1000$, as readily achievable in current experiments~\cite{rossi2024quantum, lindner_hollow-core_2024}, would enable much larger coherence lengths displayed by the gray dotted line in Fig.~\ref{fig:4}.

\paragraph{Summary and Outlook}
In summary, we have demonstrated the accelerated expansion of a nanoparticle’s motional state in a dark, inverted harmonic potential. This approach minimizes decoherence due to light scattering and blackbody radiation while enabling faster protocols. Our results  highlight the inverted potential as a powerful tool for the generation of large quantum states. We achieve state sizes comparable to the size of the nanoparticle itself, attaining an expansion factor of nearly three orders of magnitude within a few hundred microseconds. This corresponds to a squeezing level of $\mathcal{S}_z = -10\ log_{10}\left(\eta_z^{-2}\right)= 59.6$dB for a pure state in the absence of decoherence and correlations between position and momentum. 
Furthermore, our method enables the modification of the potential from purely inverted potentials to a double-well configuration~\cite{roda2024macroscopic}. The quartic contribution of the double well introduces nonlinear dynamics, enabling the study of non-Gaussian states. Both inverted and nonharmonic potentials are essential ingredients for generating macroscopic quantum superpositions.\\
\vspace{10pt}

\textbf{Acknowledgements:} This research was supported by the European Research Council (ERC) through grant Q-Xtreme ERC 2020-SyG (grant agreement number 951234).  ARC acknowledges funding from the European Union’s Horizon 2020 research and innovation programme under the Marie Skłodowska Curie grant agreement No. [101103589] (DecoXtreme, HORIZON-MSCA-2022-PF-01-01). We acknowledge valuable discussions with the Q-Xtreme synergy consortium. \\
\vspace{10pt}

%

\end{document}